\documentclass[aps,preprint]{revtex4}%
\usepackage{amsfonts}
\usepackage{amsmath}
\usepackage{amssymb}
\usepackage{graphicx}%
\providecommand{\U}[1]{\protect\rule{.1in}{.1in}}
\newtheorem{theorem}{Theorem}

\newtheorem{example}[theorem]{Example}

\begin{document}
\title{Analysis of multiple data sequences with different distributions: defining
common principal component axes by ergodic sequence generation and multiple
reweighting composition}
\author{Ikuo Fukuda$^{1}$ and Kei Moritsugu$^{2}$}
\affiliation{{\normalsize \noindent}$^{1}\footnote{Author to whom correspondence should be
addressed: ifukuda@sim.u-hyogo.ac.jp}${\normalsize Graduate School of
Simulation Studies, University of Hyogo, Kobe 650-0047, Japan}}
\affiliation{{\normalsize \noindent}$^{2}${\normalsize Graduate School of Medical Life
Science, Yokohama City University, Yokohama 230-0045, Japan}}

\begin{abstract}
Principal component analysis (PCA) defines a reduced space described by PC
axes for a given multidimensional-data sequence to capture the variations of
the data. In practice, we need multiple data sequences that accurately obey
individual probability distributions and for a fair comparison of the
sequences we need PC axes that are common for the multiple sequences but
properly capture these multiple distributions. For these requirements, we
present individual ergodic samplings for these sequences and provide special
reweighting for recovering the target distributions.

\end{abstract}
\date{December 28, 2020}
\maketitle

\section{Introduction\label{Introduction}}

Principal component analysis (PCA) is one of the statistical analysis that
defines a framework, viz., a reduced space determined by the PC axes, for a
given multidimensional-data sequence to properly capture the
varieties/variations of the data. Our target is a sequence of$\ m$dimensional
$N_{\text{step}}$\ data, $\mathcal{X}\equiv\left\{  x^{[1]},\ldots
,x^{[N_{\text{step}}]}\right\}  \subset\mathbb{R}^{m}$, generated by a
dynamical system or computer simulation such as molecular dynamics (MD) or
Monte Carlo (MC). Our purpose is, first, to generate a data sequence that
enables to completely describe a specific probability distribution $P$, which
determines the variety of the data. The Boltzmann--Gibbs (BG) distribution,
for example, is applicably useful as the physicochemical probability
distribution since it enables realistic comparisons with experiments performed
at constant temperature.\ Our second purpose concerns with two or more given
data sequences, say, $\mathcal{Y}\equiv\left\{  y^{[1]},\ldots
,y^{[N_{\text{step}}^{\prime}]}\right\}  \subset\mathbb{R}^{m}$ in addition to
$\mathcal{X}$, and is to constitute PC axes for a joint system described by
$\mathcal{X}\cup\mathcal{Y}\subset\mathbb{R}^{m}$.\ Namely, instead of seeking
independently the PC axis for $\mathcal{X}$\ and that for $\mathcal{Y}$, we
seek a common set of PC axes for both sequences $\mathcal{X}$ and
$\mathcal{Y}$, enabling us to fairly compare them in a unified framework.

Our purpose is thus (i) to generate two (or more if needed) sequences
$\mathcal{X}$\ and $\mathcal{Y}$\ that can accurately reproduce distributions
$P$\ and $P^{\prime}$, respectively; and (ii) to seek for $\mathcal{X}%
\cup\mathcal{Y}$\ unique PC axes\ that duly capture the individual varieties
for $\mathcal{X}$ and those for $\mathcal{Y}$, which are different, in
general, according to the difference between $P$\ and $P^{\prime}$. MD or
MC\ protocol has been usually used for a practical purpose to generate the BG
distribution, whereas the accurate BG distribution is not generated in general
due to the broken ergodicity and/or sampling insufficiency. We also seek a
desired distribution $P$, not limited to the BG distribution. These problems
can be overcome by an enhanced sampling method that can generate a modified
distribution $\tilde{P}$\ to recover the ergodicity, with help of reweighting
technique to reproduce $P$. Although this will be a solutions to (i), it is
far from a solution to (ii). This is because we require two different
reweightings for $P$\ and $P^{\prime}$, which will not be easily compatible
with the notion of the composition of the two data sequences. We here present
a scheme to solve both (i) and (ii) with providing common PC axes.
Furthermore, we introduce a scheme to seek (absolute continuous) distributions
$P$\ and $P^{\prime}$ on the common PCA\ space defined by the resultant PC axes.

\section{Basics}

We give a simple and probable setting of the data, as encountered in MD, which
can be generalized or transformed into other context without difficulty. We
also treat only two data sequences $\mathcal{X}$\ and $\mathcal{Y}$\ to
simplify the notations, and generalization into multiple sequences can be
simply done.

\subsection{Data sequences}

Let $x=(x_{1},\ldots,x_{n_{1}})\in\mathbb{R}^{n_{1}}$\ represent coordinates
of a given physical/dynamical system (we call it \textquotedblleft
system~1\textquotedblright) with $n_{1}$\ degrees of freedom,\ and let
\begin{equation}
\{x(\nu\Delta t)\in\mathbb{R}^{n_{1}}\mid\nu=1,\ldots,N_{\text{step}}\}
\end{equation}
a coordinate sequence, from time $\Delta t$ to time $N_{\text{step}}\Delta t$,
generated from this system. Instead of all coordinates $x(\nu\Delta
t)=(x_{1}(\nu\Delta t),\ldots,x_{n}(\nu\Delta t))$, our interest is in its $m$
parts, $(x_{k_{1}}(\nu\Delta t),\ldots,x_{k_{m}}(\nu\Delta t))=:\pi
(x(\nu\Delta t))\in\mathbb{R}^{m}$ for every time $\nu\Delta t$. Here we
denote a projection map for $x\in\mathbb{R}^{n_{1}}$\ by
\begin{equation}
\pi(x)=(\pi_{i}(x))_{i=1,\ldots,m}=(x_{k_{i}})_{i=1,\ldots,m}\in\mathbb{R}%
^{m}. \label{def of projection of system1}%
\end{equation}
We thus describe each member in $\mathcal{X=}\left\{  x^{[1]},\ldots
,x^{[N_{\text{step}}]}\right\}  \subset\mathbb{R}^{m}$ with a component
index$\ i$\ and time index $\nu$ such that\
\begin{equation}
x_{i}^{[\nu]}\equiv x_{k_{i}}(\nu\Delta t)=\pi_{i}(x(\nu\Delta t))\in
\mathbb{R},\text{ }i=1,\ldots,m. \label{target coordinate for sys1}%
\end{equation}

We also consider $\{y(\nu\Delta t)\in\mathbb{R}^{n_{2}}\mid\nu=1,\ldots
,N_{\text{step}}^{\prime}\}$, a sequence of coordinates of $n_{2}$\ degrees of
freedom,\ generated by other physical system (\textquotedblleft
system~2\textquotedblright), and are interested in $m$ parts $(y_{l_{1}}%
(\nu\Delta t),\ldots,y_{l_{m}}(\nu\Delta t))=:\pi^{\prime}(y(\nu\Delta
t))\in\mathbb{R}^{m}$ (where $\pi:\mathbb{R}^{n_{1}}\rightarrow\mathbb{R}^{m}$
and $\pi^{\prime}:\mathbb{R}^{n_{2}}\rightarrow\mathbb{R}^{m}$ are projections
into the same dimensional space $\mathbb{R}^{m}$),
\begin{equation}
y_{i}^{[\nu]}\equiv y_{l_{i}}(\nu\Delta t)=\pi_{i}^{\prime}(y(\nu\Delta
t))\in\mathbb{R},\text{ }i=1,\ldots,m. \label{target coordinate for sys2}%
\end{equation}
Hence, a sequence $\mathcal{Y}\equiv\left\{  y^{[1]},\ldots,y^{[N_{\text{step}%
}^{\prime}]}\right\}  \subset\mathbb{R}^{m}$\ arrises.

\begin{example}
$x^{[\nu]}\in\mathbb{R}^{m}$\ is e.g.,\ coordinates of$\ m/3$ C$_{\alpha}%
$-atoms of a protein consisting of $N=n/3$\ atoms, and $\mathcal{X=}\left\{
x^{[1]},\ldots,x^{[N_{\text{step}}]}\right\}  \subset\mathbb{R}^{m}$\ is a
certain sequence of $N_{\text{step}}$\ coordinates for $m$\ C$_{\alpha}$-atoms
of protein $X$. We are interested in comparison between protein $X$\ and other
protein $Y$ that may have different numbers of atoms but have the same number
of\ C$_{\alpha}$-atom coordinates, $m$, describing $y^{[\nu]}\in\mathbb{R}%
^{m}$\ and\ yielding $\mathcal{Y=}\left\{  y^{[1]},\ldots,y^{[N_{\text{step}%
}^{\prime}]}\right\}  $.
\end{example}

\subsection{PCA: review}

PCA defines a linear map from the target data space $\mathbb{R}^{m}$\ into a
reduced space, $\varphi:\mathbb{R}^{m}\rightarrow\mathbb{R}^{l}$, where
$l\equiv\dim\varphi(\mathbb{R}^{m})$ is\ less than $m\ $and typically $2$ or
$3$.\ For system~1, this map is designed so as to capture the variety of data
sequence $\mathcal{X}=\left\{  x^{[1]},\ldots,x^{[N_{\text{step}}]}\right\}
\subset\mathbb{R}^{m}$ and represent them on the reduced space $\mathbb{R}%
^{l}$. The map $\varphi$\ can be constructed via the $m\times m$\ symmetric
covariance matrix
\[
S:=\frac{1}{N_{\text{step}}}\sum_{\nu=1}^{N_{\text{step}}}\left(  x^{[\nu
]}-\bar{x}\right)  \otimes\left(  x^{[\nu]}-\bar{x}\right)  =\frac
{1}{N_{\text{step}}}\sum_{\nu=1}^{N_{\text{step}}}\left(  \left(  x_{i}%
^{[\nu]}-\bar{x}_{i}\right)  \left(  x_{j}^{[\nu]}-\bar{x}_{j}\right)
\right)  _{i,j=1\ldots,m},
\]
where $\bar{x}$\ is the average of the data,
\[
\mathbb{R}^{m}\ni\bar{x}\equiv\left(  \bar{x}_{i}\right)  _{i=1\ldots
,m}:=\frac{1}{N_{\text{step}}}\sum_{\nu=1}^{N_{\text{step}}}x^{[\nu]}=\left(
\frac{1}{N_{\text{step}}}\sum_{\nu=1}^{N_{\text{step}}}x_{i}^{[\nu]}\right)
_{i=1\ldots,m}.
\]
That is, by seeking $l$\ eigenvalues $\lambda_{1}\geq\lambda_{2}\geq\cdots
\geq\lambda_{l}$\ and the corresponding (normalized) eigenvectors $u_{1},$
$u_{2},$ $\cdots,u_{l}\in\mathbb{R}^{m}\ $for $S$, the map $\varphi$ is
defined by a projection into $\left\langle u_{1},u_{2},\cdots,u_{l}%
\right\rangle $, which is isomorphic to $\mathbb{R}^{l}$, such that
\begin{equation}
\varphi:\mathbb{R}^{m}\rightarrow\mathbb{R}^{l},x\overset{\text{\textrm{d}}%
}{\mapsto}\left(  \left(  x|u_{k}\right)  \right)  _{k=1\ldots,l}, \label{fai}%
\end{equation}
where $\left(  \cdot|\cdot\right)  $ is the inner product of $\mathbb{R}^{m}$.
Here, $u_{1}$\ is interpreted to indicate the direction to which the variety
of the data in $\mathcal{X}$ takes the maximum, $u_{2}$\ the second, and so
on. Thus the average $\bar{x}$\ and matrix $S$ are key quantities to
completely determine the PC axes. Similarly, the average and matrix are
defined for sequence $\mathcal{Y=}\left\{  y^{[1]},\ldots,y^{[N_{\text{step}%
}^{\prime}]}\right\}  $\ for\ system~2.

\section{Method for solution}

\subsection{Strategy}

Suppose that there exist \textit{ideal }time series for systems 1 and 2,
i.e.,
\begin{equation}
\{\check{x}(\nu\Delta t)\in\mathbb{R}^{n_{1}}\mid\nu=1,\ldots,N_{\text{step}%
}\} \label{exact time series of system 1}%
\end{equation}
that exactly\ obeys a distribution\ $P$ for system~1\ and
\begin{equation}
\{\check{y}(\nu\Delta t)\in\mathbb{R}^{n_{2}}\mid\nu=1,\ldots,N_{\text{step}%
}^{\prime}\} \label{exact time series of system 2,}%
\end{equation}
that exactly\ obeys a distribution\ $P^{\prime}$ for system~2 (we use
\textquotedblleft$\,\check{\cdot}\,$\textquotedblright\ to represent the
ideal), wherein$\ N_{\text{step}}$ and $N_{\text{step}}^{\prime}$\ should be
sufficiently large. This ideal situation will directly satisfy requirement~(i)
in section~\ref{Introduction}. Under this situation, requirement (ii) can be
fulfilled by constructing a PC map $\mathbb{R}^{m}\rightarrow\mathbb{R}^{l}%
$\ using a simple sum of coordinates for the two systems
\begin{equation}
\frac{1}{N_{\text{Tot}}}\left[  \sum_{\nu=1}^{N_{\text{step}}}\check{x}%
^{[\nu]}+\sum_{\nu=1}^{N_{\text{step}}^{\prime}}\check{y}^{[\nu]}\right]
\equiv\overset{\_}{\check{z}}\in\mathbb{R}^{m} \label{tot ave}%
\end{equation}
along with a simple sum of the covariance matrices for the two systems
\begin{equation}
\frac{1}{N_{\text{Tot}}}\left[  \sum_{\nu=1}^{N_{\text{step}}}\left(
\check{x}^{[\nu]}-\overset{\_}{\check{z}}\right)  \otimes\left(  \check
{x}^{[\nu]}-\overset{\_}{\check{z}}\right)  +\sum_{\nu=1}^{N_{\text{step}%
}^{\prime}}\left(  \check{y}^{[\nu]}-\overset{\_}{\check{z}}\right)
\otimes\left(  \check{y}^{[\nu]}-\overset{\_}{\check{z}}\right)  \right]
\equiv\check{T}\in\text{End}\mathbb{R}^{m}, \label{tot mat}%
\end{equation}
where $N_{\text{Tot}}\equiv N_{\text{step}}+N_{\text{step}}^{\prime}$. Here,
$\check{x}_{i}^{[\nu]}=\pi_{i}(\check{x}(\nu\Delta t))$ and $\check{y}%
_{i}^{[\nu]}=\pi_{i}^{\prime}(\check{y}(\nu\Delta t))$\ for $i=1,\ldots
,m$\ are projected coordinate components for the ideal time series,
corresponding to Eqs.~(\ref{target coordinate for sys1})
and~(\ref{target coordinate for sys2}), respectively.

\textit{Remark}--. The sum of the first and the second terms used in
Eq.~(\ref{tot ave}) and that in Eq.~(\ref{tot mat}) are mathematically well
defined because the projections ($\pi$ and $\pi^{\prime}$) are into the
identical space $\mathbb{R}^{m}$. These sums are also the most natural
expressions to represent the composed sequence $\left\{  \check{x}%
^{[1]},\ldots,\check{x}^{[N_{\text{step}}]}\right\}  \cup\left\{  \check
{y}^{[1]},\ldots,\check{y}^{[N_{\text{step}}^{\prime}]}\right\}  $. In
practice, we should also assume that these simple sums are meaningful in the
context of, e.g., chemical or physical sense. A generalization is
straightforward such that the simple sums can be replaced into weighted sums
such as $w_{1}\sum_{\nu=1}^{N_{\text{step}}}\check{x}^{[\nu]}+w_{2}\sum
_{\nu=1}^{N_{\text{step}}^{\prime}}\check{y}^{[\nu]}$ or a more general form
such as $g\left(  \sum_{\nu=1}^{N_{\text{step}}}\check{x}^{[\nu]},\sum_{\nu
=1}^{N_{\text{step}}^{\prime}}\check{y}^{[\nu]}\right)  $ using a certain
function $g$\ ($m$\ can be changed in system 2) if necessary and meaningful.

As will be discussed below, however, generation of ideal time
series~(\ref{exact time series of system 1})
and~(\ref{exact time series of system 2,})\ is nontrivial. Despite this fact,
our purpose is to have accurate $\overset{\_}{\check{z}}$\ and $\check{T}$,
which are described by the ideal time series. We will meet this seemingly
contradictory demand by deriving quantities that are equivalent\ to
Eqs.~(\ref{tot ave}) and~(\ref{tot mat}).

\subsection{Solution to requirement (i): ergodic sequence generation}

Generation of ideal time series corresponding to arbitrary distribution within
a practical $N_{\text{step}}$\ is hard in general. For example, statistics of
the data $\{x(\nu\Delta t)\in\mathbb{R}^{n}\mid\nu=1,\ldots,N_{\text{step}}\}$
generated by a conventional canonical simulation (for system~1) does not
accurately obey the BG distribution and often becomes significantly inaccurate
and uncontrollable due to broken ergodicity and/or sampling inefficiency.
Thus, we cannot meet requirements (i) and (ii) with a conventional method.
Even if an accurate sampling method exists that can directly generate any
distribution, generation of the accurate BG distribution is significantly time
consuming due to the feature of the distribution, i.e., the exponential
damping with respect to the physical system energy.

Hereafter, we assume a distribution with the form of $P=\rho(x,p)dxdp$ and set
it as the BG distribution, viz., $\rho(x,p)\varpropto\rho_{\text{BG}%
}(x,p)\equiv\rho_{\text{BG}}(x,p;\beta)\equiv\exp\left[  -\beta E(x,p)\right]
$, considering the utility and a challenge to the faced problem, although
$\rho$ can be an arbitrarily given smooth density function in principle. This
is for system~1, where $p=(p_{1},\ldots,p_{n_{1}})\in\mathbb{R}^{n_{1}}$\ and
$E(x,p)\equiv U(x)+K(p)$\ are the momenta and the total energy for system~1,
respectively ($x\in D\subset\mathbb{R}^{n_{1}}$ and $K(p)=(p|\mathbf{M}p)/2$).
It also applies to system~2, where $P^{\prime}=\rho^{\prime}%
(y,q)dydq\varpropto\rho_{\text{BG}}^{\prime}(y,q)dydq\equiv\exp\left[
-\beta^{\prime}E^{\prime}(y,q)\right]  dydq$ with $E^{\prime}(y,q)=U^{\prime
}(y)+K^{\prime}(q)$.

In our method, (i) will be fulfilled by an indirect method, which does not
mean the direct production of sequences~(\ref{exact time series of system 1})
and~(\ref{exact time series of system 2,}) but utilizes a reweighting
technique. Now, the ideal time series for a suitably defined distribution can
be generated by double density dynamics~\cite{FM1} or coupled Nos\'{e}-Hoover
(cNH) equation~\cite{FM3}. The latter realizes the equality
\begin{equation}
\bar{A}:=\lim\limits_{\tau\rightarrow\infty}\dfrac{1}{\tau}%
{\displaystyle\int_{0}^{\tau}}
A(x(t),p(t))dt=\frac{\int_{D\times\mathbb{R}^{n}}dxdpA(x,p)\rho_{\text{R}%
}(x,p)}{\int_{D\times\mathbb{R}^{n}}dxdp\rho_{\text{R}}(x,p)}=:\left\langle
A\right\rangle _{\text{R}} \label{PS variable average}%
\end{equation}
for any physical quantity $A:D\times\mathbb{R}^{n}\rightarrow\mathbb{R}$ and
any trajectory $\{(x(t),p(t))\}$\ of systems~1 under the ergodic
condition~\cite{FM3}. Although $\rho_{\text{R}}$\ is a smooth density that can
be arbitrarily designed in principle, the cNH utilized a delocalized density
\[
\rho_{\text{R}}(x,p)\equiv\int\rho_{\text{BG}}(x,p;\beta)\,f(\beta)\,d\beta,
\]
using a properly set function $f$ to efficiently cover the target region for
$\rho_{\text{BG}}$ and enhance the phase-space sampling~\cite{FM4}%
.\ Equation~(\ref{PS variable average}) enables reweighting to the target
density $\rho_{\text{BG}}$~\cite{FM3}:
\begin{align}
&  \overline{A\rho_{\text{BG}}/\rho_{\text{R}}}\left/  \overline
{\rho_{\text{BG}}/\rho_{\text{R}}}\right. \nonumber\\
&  =\left\langle A\rho_{\text{BG}}/\rho_{\text{R}}\right\rangle _{\text{R}%
}\left/  \left\langle \rho_{\text{BG}}/\rho_{\text{R}}\right\rangle
_{\text{R}}\right. \nonumber\\
&  =\int_{D\times\mathbb{R}^{n}}A(x,p)\rho_{\text{BG}}(x,p)dxdp\left/
\int_{D\times\mathbb{R}^{n}}\rho_{\text{BG}}(x,p)dxdp\right. \nonumber\\
&  =\int AdP\left/  \int dP\right.  . \label{ro_TRG integ}%
\end{align}

Thus, $\{x(\nu\Delta t)\in\mathbb{R}^{n_{1}}\mid\nu=1,\ldots,N_{\text{step}%
}\}$\ generated by the cNH satisfies (i). To see this directly, first define a
weight
\begin{equation}
w(x,p):=\frac{\rho_{\text{BG}}(x,p)/\rho_{\text{R}}(x,p)}{\sum_{\nu
=1}^{N_{\text{step}}}(\rho_{\text{BG}}/\rho_{\text{R}})(x(\nu\Delta
t),p(\nu\Delta t))}. \label{weight for system1}%
\end{equation}
Second, for any function $B:D\rightarrow\mathbb{R}$, utilize reweighting
formula~(\ref{ro_TRG integ}) to get
\begin{align}
&  \sum_{\nu=1}^{N_{\text{step}}}w(x(\nu\Delta t),p(\nu\Delta t))\text{
}B(x(\nu\Delta t))\nonumber\\
&  =\frac{\frac{1}{N_{\text{step}}}\sum_{\nu=1}^{N_{\text{step}}}B(x(\nu\Delta
t))(\rho_{\text{BG}}/\rho_{\text{R}})(x(\nu\Delta t),p(\nu\Delta t))}{\frac
{1}{N_{\text{step}}}\sum_{\nu=1}^{N_{\text{step}}}(\rho_{\text{BG}}%
/\rho_{\text{R}})(x(\nu\Delta t),p(\nu\Delta t))}\nonumber\\
&  \simeq\frac{\lim\limits_{\tau\rightarrow\infty}\dfrac{1}{\tau}%
{\displaystyle\int_{0}^{\tau}}
B(x(t))(\rho_{\text{BG}}/\rho_{\text{R}})(x(t),p(t))dt}{\lim\limits_{\tau
\rightarrow\infty}\dfrac{1}{\tau}%
{\displaystyle\int_{0}^{\tau}}
(\rho_{\text{BG}}/\rho_{\text{R}})(x(t),p(t))dt}\nonumber\\
&  =\int BdP\left/  \int dP\right.  . \label{reproduction of P}%
\end{align}
Since $B$\ is arbitrary, Eq.~(\ref{reproduction of P}) means the reproduction
of distribution $P$, indicating the satisfactions of (i) for system~1.

For systems~2, using $\rho_{\text{R}}^{\prime}(y,q)\equiv\int\rho_{\text{BG}%
}^{\prime}(y,q;\beta)\,f^{\prime}(\beta)\,d\beta$ and
\[
w^{\prime}(y,q):=\frac{\rho_{\text{BG}}^{\prime}(y,q)/\rho_{\text{R}}^{\prime
}(y,q)}{\sum_{\nu=1}^{N_{\text{step}}^{\prime}}(\rho_{\text{BG}}^{\prime}%
/\rho_{\text{R}}^{\prime})(y(\nu\Delta t),q(\nu\Delta t))}%
\]
yields
\begin{align}
&  \sum_{\nu=1}^{N_{\text{step}}^{\prime}}w^{\prime}(y(\nu\Delta
t),q(\nu\Delta t))\text{ }B^{\prime}(y(\nu\Delta t))\nonumber\\
&  \simeq\frac{\lim\limits_{\tau\rightarrow\infty}\dfrac{1}{\tau}%
{\displaystyle\int_{0}^{\tau}}
B^{\prime}(y(t))(\rho_{\text{BG}}^{\prime}/\rho_{\text{R}}^{\prime
})(y(t),q(t))dt}{\lim\limits_{\tau\rightarrow\infty}\dfrac{1}{\tau}%
{\displaystyle\int_{0}^{\tau}}
(\rho_{\text{BG}}^{\prime}/\rho_{\text{R}}^{\prime})(y(t),q(t))dt}\nonumber\\
&  =\int B^{\prime}dP^{\prime}\left/  \int dP^{\prime}.\right.
\label{reproduction of P'}%
\end{align}
Combining the results for the two systems manifests the satisfaction of (i).
Note that, a method, instead of the cNH, suffices if it ensures
Eq.~(\ref{PS variable average}) for any $A$ and any trajectory and if the
relationship~(\ref{ro_TRG integ}) works for the target BG\ distribution, for
any systems.

\subsection{Solution to requirement (ii): multiple reweighting composition}

Based on the results obtained above, requirement (ii) can be satisfied as
follows. By substituting $B\equiv\pi_{i}$\ in Eq.~(\ref{reproduction of P}%
)\ and $B^{\prime}\equiv\pi_{i}^{\prime}$ in Eq.~(\ref{reproduction of P'}),
we have \
\begin{align}
\bar{z}_{i}^{W}  &  :=\frac{N_{\text{step}}}{N_{\text{Tot}}}\sum_{\nu
=1}^{N_{\text{step}}}w(x(\nu\Delta t),p(\nu\Delta t))x_{i}^{[\nu]}%
+\frac{N_{\text{step}}^{\prime}}{N_{\text{Tot}}}\sum_{\nu=1}^{N_{\text{step}%
}^{\prime}}w^{\prime}(y(\nu\Delta t),q(\nu\Delta t))y_{i}^{[\nu]}\nonumber\\
&  \simeq\frac{N_{\text{step}}}{N_{\text{Tot}}}\int\pi_{i}(x)dP\left/  \int
dP\right.  +\frac{N_{\text{step}}^{\prime}}{N_{\text{Tot}}}\int\pi_{i}%
^{\prime}(y)dP^{\prime}\left/  \int dP^{\prime}\right. \nonumber\\
&  \simeq\frac{N_{\text{step}}}{N_{\text{Tot}}}\frac{1}{N_{\text{step}}}%
\sum_{\nu=1}^{N_{\text{step}}}\check{x}_{i}^{[\nu]}+\frac{N_{\text{step}%
}^{\prime}}{N_{\text{Tot}}}\frac{1}{N_{\text{step}}^{\prime}}\sum_{\nu
=1}^{N_{\text{step}}^{\prime}}\check{y}_{i}^{[\nu]}\nonumber\\
&  =\overset{\_}{\check{z}}_{i}\text{ for }i=1,\ldots,m, \label{zi approx}%
\end{align}
where the third line comes from the fact that the ideal time
series~(\ref{exact time series of system 1})
and~(\ref{exact time series of system 2,}) obey the distributions $P$\ and
$P^{\prime}$, respectively. Consequently, the target quantity,
Eq.~(\ref{tot ave}), is obtained by calculating $\bar{z}_{i}^{W}$. We also
have
\begin{align}
T_{ij}^{W}  &  :=\frac{N_{\text{step}}}{N_{\text{Tot}}}\sum_{\nu
=1}^{N_{\text{step}}}w(x(\nu\Delta t),p(\nu\Delta t))(x_{i}^{[\nu]}-\bar
{z}_{i}^{W})(x_{j}^{[\nu]}-\bar{z}_{j}^{W})\nonumber\\
&  \text{ \ }+\frac{N_{\text{step}}^{\prime}}{N_{\text{Tot}}}\sum_{\nu
=1}^{N_{\text{step}}^{\prime}}w^{\prime}(y(\nu\Delta t),q(\nu\Delta
t))(y_{i}^{[\nu]}-\bar{z}_{i}^{W})(y_{j}^{[\nu]}-\bar{z}_{j}^{W})\nonumber\\
&  \simeq\frac{N_{\text{step}}}{N_{\text{Tot}}}\frac{\int(x_{k_{i}}-\bar
{z}_{i}^{W})(x_{k_{j}}-\bar{z}_{j}^{W})dP}{\int dP}+\frac{N_{\text{step}%
}^{\prime}}{N_{\text{Tot}}}\frac{\int\text{ }(y_{l_{i}}-\bar{z}_{i}%
^{W})(y_{l_{j}}-\bar{z}_{j}^{W})dP^{\prime}}{\int dP^{\prime}}\nonumber\\
&  \simeq\frac{1}{N_{\text{Tot}}}\sum_{\nu=1}^{N_{\text{step}}}\left(
\check{x}_{i}^{[\nu]}-\bar{z}_{i}^{W}\right)  \left(  \check{x}_{j}^{[\nu
]}-\bar{z}_{j}^{W}\right)  +\frac{1}{N_{\text{Tot}}}\sum_{\nu=1}%
^{N_{\text{step}}^{\prime}}(\check{y}_{i}^{[\nu]}-\bar{z}_{i}^{W})(\check
{y}_{j}^{[\nu]}-\bar{z}_{j}^{W})\nonumber\\
&  \simeq\frac{1}{N_{\text{Tot}}}\sum_{\nu=1}^{N_{\text{step}}}\left(
\check{x}_{i}^{[\nu]}-\overset{\_}{\check{z}}_{i}\right)  \left(  \check
{x}_{j}^{[\nu]}-\overset{\_}{\check{z}}_{j}\right)  +\frac{1}{N_{\text{Tot}}%
}\sum_{\nu=1}^{N_{\text{step}}^{\prime}}(\check{y}_{i}^{[\nu]}-\overset
{\_}{\check{z}}_{i})(\check{y}_{j}^{[\nu]}-\overset{\_}{\check{z}}%
_{j})\nonumber\\
&  =\check{T}_{ij}\text{ for }i,j=1,\ldots,m, \label{Tij approx}%
\end{align}
using $\overset{\_}{\check{z}}_{i}\simeq\bar{z}_{i}^{W}$\ ($i=1,\ldots,m$)
concluded in Eq.~(\ref{zi approx}). Therefore, these procedures for obtaining
$\overset{\_}{\check{z}}$\ and $\check{T}$ by calculating $\bar{z}^{W}$\ and
$T^{W}$ ensure the satisfaction of (ii).

\subsection{BG distribution on the PCA space}

Hence, we have a PCA space $\mathbb{R}^{l}$ defined by map~(\ref{fai}),
$\varphi\equiv\varphi^{W}$, constructed from covariance matrix $T^{W}%
$\ obtained above. The BG distribution on the PCA space $\mathbb{R}^{l}$\ for
system~1 is formulated as an induced probability measure of $P$ on
$\mathbb{R}^{2n_{1}}$\ via a map $\varphi_{\pi}:D\times\mathbb{R}^{n_{1}%
}\rightarrow\mathbb{R}^{l},(x,p)\overset{\mathrm{d}}{{\mapsto}}\varphi
(\pi(x))$, where $\pi$\ is projection~(\ref{def of projection of system1}).
That is,
\begin{align}
P_{\varphi_{\pi}}:\mathbb{R}^{l}\supset B  &  \overset{\mathrm{d}}{{\mapsto}%
}P(\varphi_{\pi}^{-1}(B))\nonumber\\
&  =P\left(  (\varphi\circ\pi)^{-1}(B)\times\mathbb{R}^{n_{1}}\right)
\nonumber\\
&  =\frac{\int_{(\varphi\circ\pi)^{-1}(B)\times\mathbb{R}^{n_{1}}}%
\rho_{\text{BG}}(x,p)dxdp}{\int_{D\times\mathbb{R}^{n_{1}}}\rho_{\text{BG}%
}(x,p)dxdp}. \label{BG distribution on the PCA space}%
\end{align}
Here, $P_{\varphi_{\pi}}(B)$ represented by the RHS of
Eq.~(\ref{BG distribution on the PCA space}) can be evaluated for any
$B\in\mathcal{B}^{l}$, using the weight $w$ defined by
Eq.~(\ref{weight for system1}), as follows:
\begin{align}
&  \sum_{\substack{\nu=1\\\left(  \left(  x^{[\nu]}|u_{1}\right)
,\ldots\left(  x^{[\nu]}|u_{l}\right)  \right)  \in B}}^{N_{\text{step}}%
}w(x(\nu\Delta t),p(\nu\Delta t))\nonumber\\
&  \simeq\overline{\hat{\chi}_{B}\rho_{\text{BG}}/\rho_{\text{R}}}\left/
\overline{\rho_{\text{BG}}/\rho_{\text{R}}}\right. \nonumber\\
&  =\frac{\int_{D\times\mathbb{R}^{n_{1}}}\hat{\chi}_{B}(x)\rho_{\text{BG}%
}(x,p)dxdp}{\int_{D\times\mathbb{R}^{n_{1}}}\rho_{\text{BG}}(x,p)dxdp}%
\nonumber\\
&  =\frac{\int_{(\varphi\circ\pi)^{-1}(B)\times\mathbb{R}^{n_{1}}}%
\rho_{\text{BG}}(x,p)dxdp}{\int_{D\times\mathbb{R}^{n_{1}}}\rho_{\text{BG}%
}(x,p)dxdp}, \label{evaluation induced meas}%
\end{align}
where $\hat{\chi}_{B}$ is a characteristic function defined as
\[
\hat{\chi}_{B}:D\rightarrow\mathbb{R},x\overset{\text{\textrm{d}}}{\mapsto
}\left\{
\begin{array}
[c]{c}%
1\text{ \ if }\varphi(\pi(x))\in B\\
0\text{ \ otherwise \ \ }%
\end{array}
\right.  .
\]
The sum in the LHS of Eq.~(\ref{evaluation induced meas}) means that the
weight is counted if the $l$\ PC-coordinates of $x^{[\nu]}$\ fall into the bin
$B$. These results for system~1 similarly apply to system 2.

\section{Numerics}%

\begin{figure}
[ptb]
\begin{center}
\includegraphics[
height=2.8154in,
width=6.3332in
]%
{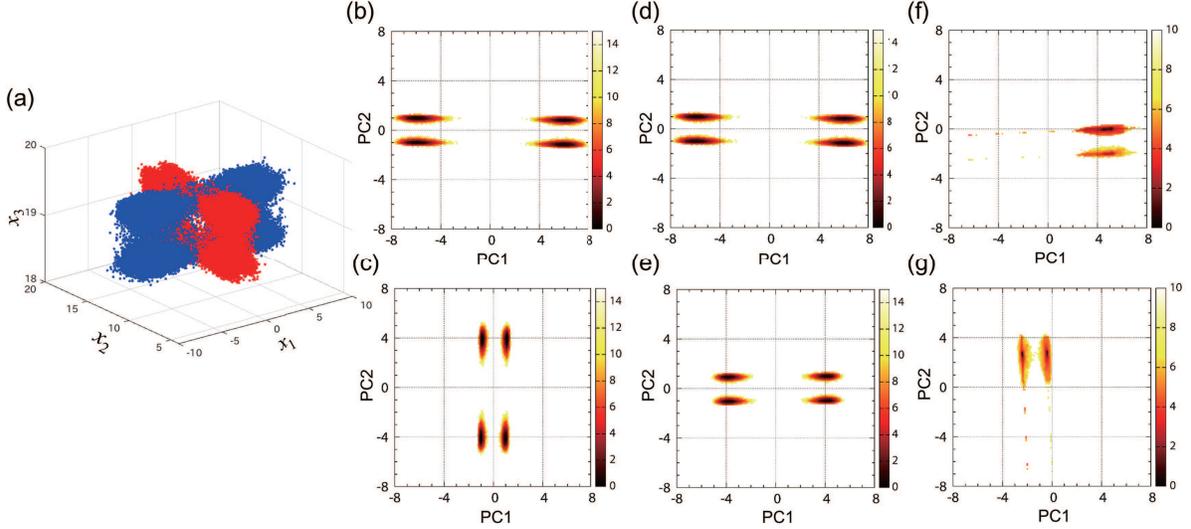}%
\caption{(a) Scatter plots for sequence $\mathcal{X}\cup\mathcal{Y}$, where
$\mathcal{X}$\ was\ obtained by cNH simulation of system~1 (blue) with
$n_{1}=4$ and by projection $\pi:(x_{1},x_{2},x_{3},x_{4})\mapsto(x_{1}%
,x_{2},x_{3})$, while $\mathcal{Y}$\ was that of system~2 (red) with $n_{2}%
=4$\ and $\pi^{\prime}=\pi$; (b) and (c) show current PCA results for
systems~1\ and 2, respectively, obtained by the common PC axes, with
BG\ distributions shown by color; (d) is PCA result with BG\ distribution for
system~1\ by $\mathcal{X}$, and (e) is that for system~2\ by $\mathcal{Y}$;
(f) and (g) are results\ on $\mathcal{X}_{\text{c}}\cup\mathcal{Y}_{\text{c}}$
obtained by canonical simulations for systems~1 and 2, respectively.}%
\end{center}
\end{figure}

To illustrate our method, it has been applied to \textquotedblleft
system~1\textquotedblright\ and \textquotedblleft system~2\textquotedblright%
\ modeled with four degrees of freedom ($n_{1}=n_{2}=4$) described by
potential function $U(x)=\sum_{i=1}^{4}\frac{10}{b_{i}^{4}}\left(  \left(
x_{i}-d_{i}\right)  ^{2}-b_{i}^{2}\right)  ^{2}+\sum_{i=1}^{3}\frac{k}%
{2}\left(  x_{i}-x_{i+1}-d_{i}+d_{i+1}\right)  ^{2}$. The difference between
the two systems is only in the values of "intra" parameters $b_{1}$\ and
$b_{2}$\ ($b_{1}=6,b_{2}=1$ for system~1; $b_{1}=1,b_{2}=4$ for system~2; and
$b_{3}=b_{4}=0.4,d_{1}=0,d_{2}=12,d_{3}=19,d_{4}=21,k=10^{-5}$ for both
systems). Figure 1(a) shows plots for the sequence $\mathcal{X}\cup
\mathcal{Y}\subset\mathbb{R}^{3}$, where $\mathcal{X=}\left\{  x^{[1]}%
,\ldots,x^{[N_{\text{step}}]}\right\}  \subset\mathbb{R}^{3}$\ was\ obtained
by a cNH simulation (detailed in~\cite{FM3}) of system~1 (blue) along with a
projection $\pi:\mathbb{R}^{4}\rightarrow\mathbb{R}^{3},x\mapsto(x_{1}%
,x_{2},x_{3})$, and similarly $\mathcal{Y=}\left\{  y^{[1]},\ldots
,y^{[N_{\text{step}}]}\right\}  $\ was that for system~2 (red) with
$\pi^{\prime}=\pi$ (viz., $m=3$). The accuracies were evaluated by marginal
distributions of the reweighted BG\ distributions, where the errors from the
exact values in 2-dim distributions for major variables$\ (x_{1},x_{2})$ were
$5.6\times10^{-5}$\ and $1.9\times10^{-5}$\ (with s.d. $5.5\times10^{-3}$ and
$9.4\times10^{-3}$), which are sufficiently small~\cite{FM3}, for
systems~1\ and~2, respectively.

Figures 1(b) and 1(c) show the current PCA results with $l=2$\ for
systems~1\ and~2, respectively, which were obtained by the unique common PC
axes determined by Eqs.~(\ref{zi approx})\ and~(\ref{Tij approx}) and by
reconstructing the BG distribution via Eq.~(\ref{evaluation induced meas}).
The current method properly describes the difference between the two systems.
This is because the raw data (Fig.~1(a)) suggest the role conversion between
the first and the second degrees of freedom (i.e., system~1 has the largest
variations for $x_{1}$ and the second largest variations for $x_{2}$, while
system~2 has the largest for $y_{2}$ and the second largest for $y_{1}$), and
because the current PCA results capture the role conversion between the two
degrees of freedom via PC1 and PC2, as clearly seen by the difference between
Figs.~1(b) and 1(c), owing to the fact that PC1 and PC2 axes are common for
the two systems.

In contrast, individual procedures without data jointing by $\mathcal{X}%
\cup\mathcal{Y}$, i.e., PCA for system~1\ by $\mathcal{X}$\ and independent
PCA for system~2 by $\mathcal{Y}$ resulted in misleading results, as shown in
Figs.~1(d) and 1(e), respectively. Namely, these individual PCA results
conclude that the two systems are similar. Although such judgment whether the
PCA results are reasonable or misleading is possible in these simple model
systems, it is not for general systems. Thus, the conventional methods using
independent PCA procedures for multiple systems may lost the important
information of the original systems and lead to incorrect conclusions. Hence,
it is critical to meet requirement (ii), which is to seek for $\mathcal{X}%
\cup\mathcal{Y}$\ unique PC axes\ that duly capture system~1 with
distributions $P$\ and system~2 with $P^{\prime}$.

Figures 1(f) and 1(g) show the PCA results utilizing $\mathcal{X}_{\text{c}%
}\cup\mathcal{Y}_{\text{c}}$ composed by two conventional canonical MD
simulation output sequences $\mathcal{X}_{\text{c}}$ and $\mathcal{Y}%
_{\text{c}}$\ for systems~1 and 2, respectively. The results show less
accuracy due to the sampling inefficiency with local traps. Thus,
requirement~(i) is also critical to get the proper information of the systems.
Therefore, satisfaction for both requirements (i) and (ii) is a key to succeed
PCA to capture the difference/similarity of multiple systems. The current
method both\ satisfies.

\end{document}